\documentclass[aps,prc,twocolumn,showpacs,showkeys,superscriptaddress]{revtex4-1}
\usepackage{longtable}
\usepackage{graphicx}
\usepackage{dcolumn}
\begin{document}
%\title{Population of Neutron Rich Nuclei in A $\sim$ 150 Region: Role of p-Induced Reaction}
\title{Role of p-induced population of medium mass(A $\sim$ 150) neutron rich nuclei}
\author{D.~Banerjee}
\affiliation{Accelerator Chemistry Section, RCD(BARC), Variable Energy Cyclotron Centre, Kolkata - 700 064}
\author{A.~Saha}
\affiliation{Variable Energy Cyclotron Centre, Kolkata - 700 064, India}
\author{T.~Bhattacharjee}
\thanks{Corresponding author}
\email{btumpa@vecc.gov.in}
\affiliation{Variable Energy Cyclotron Centre, Kolkata - 700 064, India}
\author{R.~Guin}
\affiliation{Accelerator Chemistry Section, RCD(BARC), Variable Energy Cyclotron Centre, Kolkata - 700 064}
\author{S.~K.~Das}
\affiliation{Accelerator Chemistry Section, RCD(BARC), Variable Energy Cyclotron Centre, Kolkata - 700 064}
\author{ P.~Das}
\affiliation{Variable Energy Cyclotron Centre, Kolkata - 700 064, India}
\author{ Deepak Pandit}
\affiliation{Variable Energy Cyclotron Centre, Kolkata - 700 064, India}
\author{A.~Mukherjee}
\affiliation{Saha Institute of Nuclear Physics, Kolkata - 700 064}
\author{A.~Chowdhury}
\affiliation{Variable Energy Cyclotron Centre, Kolkata - 700 064, India}
\author{Soumik Bhattacharya}
\affiliation{Variable Energy Cyclotron Centre, Kolkata - 700 064, India}
\author{S.~Das Gupta}
\altaffiliation{Present address: Saha Institute of Nuclear Physics, Kolkata - 700 064}
\affiliation{Variable Energy Cyclotron Centre, Kolkata - 700 064, India}
\author{ S.~Bhattacharyya}
\affiliation{Variable Energy Cyclotron Centre, Kolkata - 700 064, India}
\author{ P.~Mukhopadhyay}
\affiliation{Variable Energy Cyclotron Centre, Kolkata - 700 064, India}
\author{ S.~R.~Banerjee}
\affiliation{Variable Energy Cyclotron Centre, Kolkata - 700 064, India}

\date{\today}

\begin{abstract}
%\noindent

Excitation functions were measured by stacked-foil activation technique for the $^{150}$Nd(p, xpyn) reaction using 97.65$\%$ enriched $^{150}$Nd target. Measurement up to $\sim$50$\%$ above barrier and down to 18$\%$ below the barrier was performed using proton beam energy (E$_p$) of 7 - 15 MeV from VECC Cyclotron. The yield of suitable $\gamma$ rays emitted following the decay of relevant evaporation residues was determined using a 50$\%$ High Purity Germanium (HPGe) detector.(p,n) cross section was found to follow the expected trend with a maximum value of 63.7(4.9)mb at E$_p$ $\sim$ 8.6 MeV. (p,2n) cross section gradually increased with E$_p$ and had maximum contribution to the total reaction cross section after E$_p$ $\sim$ 9.0 MeV. (p, p$^{\prime}$n) reaction channel also showed a reasonable yield with a threshold of  E$_p$ $\sim$ 12.0 MeV. The experimental data were corroborated with statistical model calculations using different codes, viz., CASCADE, ALICE/91 and EMPIRE3.1. All the calculations using a suitable set of global parameters could reproduce the excitation function fairly well in the present energy range.

\end{abstract}

\pacs{25.60.Dz; 25.60.Pj; 23.20.Lv; 29.40.Wk; 24.10.Ht}

\keywords{Reaction cross section; Fusion reaction; Decay gamma spectroscopy; Semiconductor detector; Statistical model calculations.}

\maketitle

\section{Introduction}
\label{intro}

The light ion induced reactions with neutron-rich targets can be considered as one of the possible avenues to study the nuclei in the neutron rich side of N-Z chart. The population of the neutron-rich nuclei in A $\sim$ 150 region with sizable yield is difficult by either fission reaction, even with trans-uranium elements, or by fusion-evaporation reaction with stable heavy-ion beams~\cite{fission1,fission2}. The information on the light-ion induced reaction cross-section (CS) is required to perform the spectroscopy of the low-lying states of these nuclei which has a lot of current physics interests~\cite{review-akjain,review-cejnar}.
%One of the important aspects is the presence of long lived $\beta$-decaying isomers in odd-odd isotopes of this mass region~\cite{review-akjain}. The information on the reaction CS can be used directly in the identification of these isomers~\cite{pm148}.
One of the important aspects is the presence of long lived $\beta$-decaying isomers, in odd-odd isotopes of this mass region, which can be identified directly from the information on the reaction CS~\cite{pm148}.
The proton induced fusion evaporation reactions in this mass region are dominated by few neutron evaporation channels with a small contribution to the single neutron and particle evaporation channels. However, it is very important to estimate the low CS of the residues produced from the particle evaporation channels, which can be used in meaningful experiments with a suitable tagging device coupled to a high efficiency $\gamma$ array~\cite{tag-expt,Nd150}. Several of these experiments in A $\sim$ 150 region are being explored at VECC, Kolkata~\cite{dae-pm,dae-sm}.

The p induced reactions on stable targets can be used as the surrogate reaction~\cite{cramer} for calculating the n capture CS for the unstable neutron rich nuclei. This has immense importance in the field of basic as well as applied nuclear physics, {\it viz.}, nuclear astrophysics, nuclear reactor technology, etc~\cite{review-surrogate}. In the universe, the  synthesis of nuclei A $\sim$ 150 region mainly takes place via slow and rapid neutron capture processes. Hence the knowledge on the CS for the n-capture reactions are required for understanding the nucleosynthesis path as well as the abundance of the nuclei. However, these reaction CS are difficult to measure mainly because of the difficulty in preparing the neutron rich radioactive targets with very low half lives. Hence, the p induced surrogate reactions can provide very important inputs to the stellar model calculations where the n capture CS is calculated with the information on the compound nucleus formation CS, obtained by measuring that from the surrogate reactions. This is more important in the nuclei which are at or near the branch point having a long half lives for nuclear $\beta$ decay. The nuclei produced as residues in the proton induced reaction on $^{150}$Nd lie near the branch point nucleus $^{148}$Pm. There is almost no data for the light ion induced reactions in this mass region, when explored in the recent compiled databases~\cite{endf}. Hence, the study of CS for the p-induced reaction on the neutron-rich targets in A$\sim$ 150 region is important and the said study can also provide inputs to the different statistical model calculations.

The measurement of the p-induced reaction CS with the Nd target has important additional interest also. The decay of $^{150}$Nd nucleus, one of the very promising candidates for neutrinoless double beta decay~\cite{dbd-Nd}, is being studied in several important underground experimental facilities~\cite{sno, dcba, supernemo}. In these experiments, Nd is either loaded in the scintillator material or is used as foils within the time projection chamber. During transportation of the Nd material to the underground laboratories, it is exposed to the Cosmic background and produces several long lived isotopes from the light charged particle induced reactions~\cite{guesser}. These isotopes generate the possible backgrounds in the $^{150}$Nd double beta decay experiments with very low real yields~\cite{nemo3-expt}. A measure for the majority of this background can be obtained from the CS measurement of the proton induced reactions on different Nd isotopes as proton constitutes $\sim$ 90$\%$ of the total Cosmic particles hitting the earth's surface.

In the present work, the excitation functions of the (p, xnyp) reactions have been measured using stacked-foil activation technique~\cite{ismail} with the 97.65$\%$ enriched $^{150}$Nd target. A similar work was done by O. Lebeda {\it et al.,}~\cite{lebeda-1} using $^{nat}$Nd target.
A preliminary result of our measurement with enriched target was reported in ref.~\cite{dae-12}. Following our preliminary report, another set of measurement has been reported by O. Lebeda {\it et al.,}~\cite{lebeda-2} which also used $^{nat}$Nd target.
The use of enriched Nd target in our work has facilitated the measurement of absolute CS for the p-induced reaction on $^{150}$Nd for the first time. The measurement also provides a completely new set of data on the excitation function for the present reaction with a large number of data points in the proton energy range of 7-15 MeV.
Considering the Coulomb barrier for the present system to be $\sim$ 8.5 MeV, the fusion probability below the threshold by 1.5 MeV has been determined in the present work. The measurement of CS for $^{150}$Nd(p,n)$^{150}$Pm, $^{150}$Nd(p,2n)$^{149}$Pm and $^{150}$Nd(p,p$^{\prime}$n)$^{149}$Nd / $^{150}$Nd(p,d)$^{149}$Nd reactions have been measured giving rise to a total fusion CS for the p+$^{150}$Nd reaction. The experimental results have been corroboarated with statistical model calculation using different codes, {\it viz.,} CASCADE~\cite{cascade}, ALICE~\cite{alice} and EMPIRE3.1~\cite{empire}.

\section{Experimental Details}
\label{expt}
\begin{figure}
  \begin{center}
  %\vskip 0.5cm
  \hskip -0.5cm
  \includegraphics[height=0.7\columnwidth]{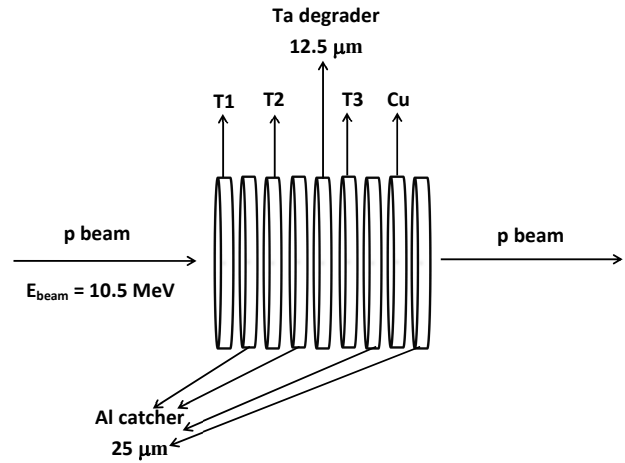}
  %\vskip -2.0cm
  \caption{The configuration of a typical stack is represented. The number, shape and size of the targets, foils and degraders were different in different stacks as detailed in the text. T1, T2 and T3 are the targets and Cu represents the monitor foil.}
  \label{fig1-stack}
  \end{center}
\end{figure}

The $^{150}$Nd (p, xnyp) reaction was carried out using 7 to 15 MeV proton beams provided by K = 130 AVF cyclotron at Variable Energy Cyclotron Centre, Kolkata. The $^{150}$Nd target was prepared  by electro-deposition technique, starting from commercially available 97.65$\%$ enriched powdered oxide sample(Nd$_2$O$_3$), on a 0.3 mil thick Aluminium (Al) foil. The Nd:O atom ratio in the deposited target was estimated by neutron activation of both the powdered material and the prepared target. The neutron beam of flux $\sim$ 1.8 x 10$^{14}$ neutrons /cm$^2$/sec was available from the research reactor facility `DHRUVA' at Bhabha Atomic Research Centre, Mumbai. The Nd:O ratio was found to be 1.6:3 in the electro-deposited target compared to 2:3 in the powdered sample. This atom ratio of 1.6:3 was used in the subsequent calculation. The isotopic impurities in the target material consisted of 0.50$\%$ $^{142}$Nd, 0.31$\%$ $^{143}$Nd, 0.68$\%$ $^{144}$Nd, 0.23$\%$ $^{145}$Nd, 0.47$\%$ $^{146}$Nd and 0.26$\%$ $^{148}$Nd, as per the data sheet provided by the supplier.
The thickness of the targets used in the experiment was within the range of 650 $\mu$g/cm$^2$ to 900 $\mu$g/cm$^2$ as determined by accurate weight difference method using analytical balance. The CS were measured by using the stacked-foil activation technique where several target stacks were irradiated with the proton beam.
Each stack contained Copper (25$\mu$m), Al (25$\mu$m) and Tantalum (12.5$\mu$m) as the beam flux monitor, catcher and degrader respectively. The number of targets in each stack was limited to a maximum value of four. A typical stack with its components has been shown in FIG.~\ref{fig1-stack}. The target stacks were prepared in such a way that every stack provides one proton energy of irradiation common to that with the previous stack. Single targets along with a monitor foil and the catchers were also irradiated at few proton energies throughout the range of 7 to 15 MeV.
%Five stacks were irradiated with proton beam of energy 8.5, 9.5, 10.5, 12.0 and 15 MeV respectively.
The beam spot on the target was confined to 6.0mm by using Al collimator in front of the target. The energy of the incident beam was determined from the knowledge of magnetic field, extraction radius and charge state of the accelerated ion. The maximum uncertainty in proton beam energy is known to be $\sim$100 keV in the present energy range. The degradation of the beam energy as a function of the depth of the stack was calculated with the help of the `Physical Calculator' available inside the code `LISE++'~\cite{lise} giving the beam energy and the target thickness as the input parameters. The beam intensity for each stack was calculated by using the known excitation function for $^{nat}$ Cu(p,xn) $^{63,65}$Zn reactions.
The absolute yields of the delayed $\gamma$ rays corresponding to the decay of the individual evaporation residues were used to determine the excitation function of the p-induced reaction on $^{150}$Nd. Following this method, both the target and the monitor foils were counted along with their respective catcher foils on a 50$\%$ HPGe detector after allowing a cooling time of 1-2 h from the end of bombardment (EOB).
The irradiated foils were placed at an appropriate distance from the detector to maintain a dead time of $\le$ 10$\%$. This was further supported by the absence of any sum peak in the $\gamma$ spectrum. The activities obtained from the stack irradiations were initially counted over a period of $\sim$10 h, considering three half lives ($\tau _{\frac{1}{2}}$) of the ground state of $^{150}$Pm ($\tau _{\frac{1}{2}}$ = 2.68 h). Each counting was performed for 10 min at an interval of $\sim$ 30 min. The counting for $^{150}$Pm could also provide the data for $^{149}$Nd ($\tau _{\frac{1}{2}}$ = 1.73 h) because of their similar $\tau _{\frac{1}{2}}$ values.
For $^{149}$Pm ($\tau _{\frac{1}{2}}$ = 53.08 h), a similar sequence of counting was done over a period of one week where the individual counting had a duration of 30 min to 2 h at an interval of one day.
The counting for $^{150}$Pm and $^{149}$Pm were done keeping the irradiated foils at a distance of 15 cm and 7 cm from the detector respectively. The absolute efficiency of the detector at these two positions was estimated by using the standard $^{152}$Eu and $^{133}$Ba sources with known activity. The energy calibration of the detector was also performed using the same set of sources. The data were acquired and analysed using a Canberra digital data processing system GENIE 2000.

\section{Data Analysis and Results}
\label{da}

The CS values ($\sigma$) of the different evaporation channels produced from the p+$^{150}$Nd reaction were estimated from the knowledge of the activity (A$_0$) of the respective evaporation residue at EOB, number of target atoms (N$_0$) and proton beam flux ($\phi$) by using the following equation:
 \begin{equation}\label{eqn-cs}
   \sigma = \frac {A_0}{N. \phi . [1 - exp({\frac{-0.693t_{irr}}{\tau _{\frac{1}{2}}}})]}\\
   \nonumber
 \end{equation}
The term in the parenthesis is known as the saturation factor used for correcting the activity lost during time of irradiation (t$_{irr}$).
A$_0$ is calculated from the yield(N$_0$) of the characteristic $\gamma$ ray at EOB, emitted following the decay of any particular evaporation residue, by using the relation $A_0 = N_0 / \epsilon \eta$,
where, $\epsilon$ is the efficiency of the detector and $\eta$ is the abundance for that particular $\gamma$ ray.
The beam flux($\phi$) was calculated using the same equation~\ref{eqn-cs} and taking the CS values for the reaction $^{nat}$Cu(p,xn)$^{65}$Zn from literature~\cite{cu-ref}. The errors in N, $N_0$ and $\epsilon$($\le$1.5$\%$, $\le$2$\%$ and $\le$5$\%$ respectively as discussed below) were taken in calculating the error in $\phi$. The error in N arises from the uncertainty in the thickness measurement of the target foil and has been considered to be $\le$1.5$\%$.

\begin{figure}
  \begin{center}
  \vskip 0.5cm
  \hskip -1.0cm
  \includegraphics[height=0.85\columnwidth, angle=-90]{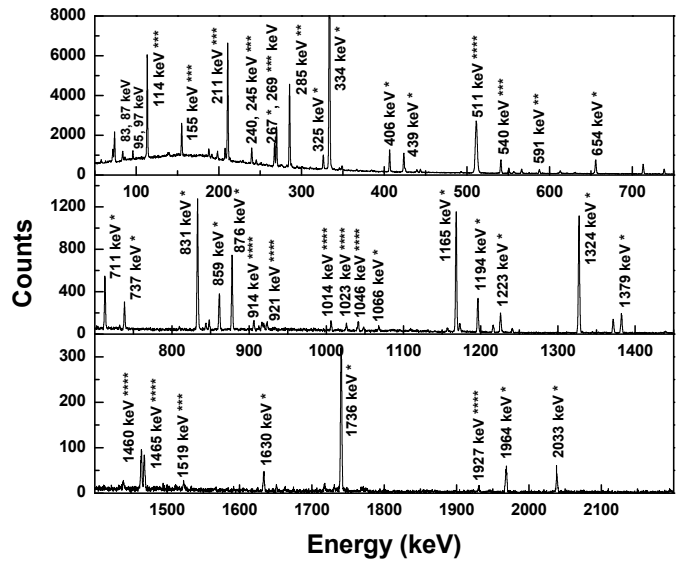}
  \vskip -1.5cm
  \caption{The total gamma spectrum taken with the 50$\%$ HPGe detector for the irradiated $^{150}$Nd target with 15 MeV proton beam. The photopeaks marked with *, **, ***, **** belong to the decay of $^{150}$Pm, $^{149}$Pm, $^{149}$Nd and the Al related or background $\gamma$ rays respectively.}
  \label{fig2-totalspec}
  \end{center}
\end{figure}

\begin{figure}
  \begin{center}
  \vskip 0.5cm
  \hskip -1.0cm
  \includegraphics[height=0.9\columnwidth, angle=-90]{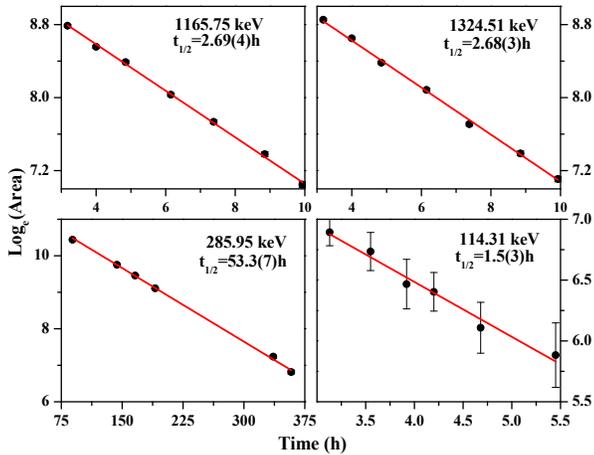}
  %\vskip -1.5cm
  \caption{The decay plots for the transitions characterising the decay of different evaporation residues formed at a particular beam energy as described in the text.}
  \label{fig3-decay}
  \end{center}
\end{figure}

\begin{figure}
  \begin{center}
  %\vskip 0.5cm
  %\hskip -2.5cm
  \includegraphics[height=\columnwidth, angle=-90]{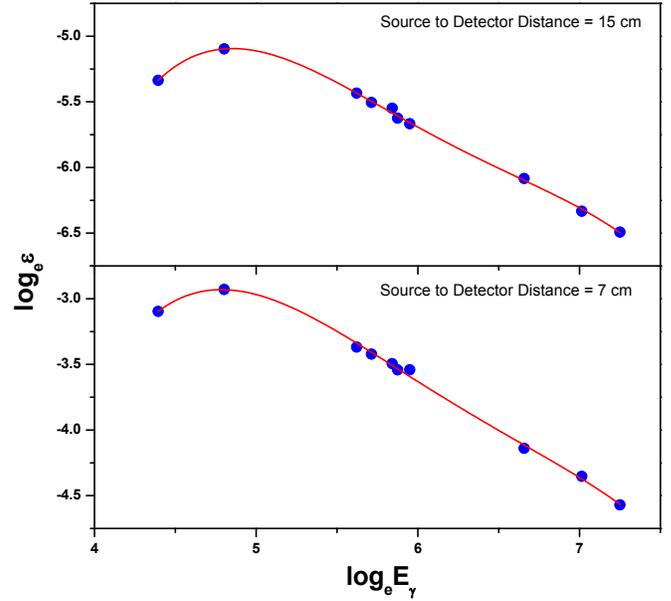}
  %\vskip -2.0cm
  \caption{The extracted efficiency of the 50$\%$ HPGe detector used in the experiment. The efficiency of the detector at a distance of 7 cm and 15 cm from the detector has been shown in the bottom panel and top panel respectively with black square and circles. The solid and dotted lines in the two figures represent the fitted curve.}
  \label{fig4-eff}
  \end{center}
\end{figure}

\begin{figure}
\begin{center}
%\vskip 3.5cm
\hskip -1.0cm
\includegraphics[height=\columnwidth, angle=-90]{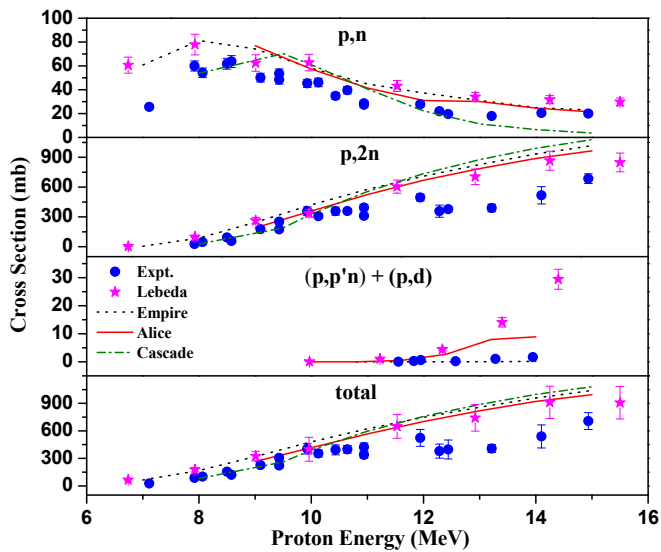}
%\vskip 1.5cm
\caption{(Colour online)Excitation function for $^{150}$Nd(p,xnyp) reaction. Experimental data points for (a)$^{150}$Pm, (b)$^{149}$Pm, (c) $^{149}$Nd and (d) total CS are shown with filled $\circ$. The calculated values of CS obtained from the work of O. Lebeda {\it et al.,} are shown with $\star$s. Theoretical CS calculated from Empire3.1, ALICE/91 and CASCADE are shown with dotted, solid and dash dotted lines respectively.}
\label{fig5-cs}
\end{center}
\end{figure}

\begin{figure}
  \begin{center}
  %\vskip 3.5cm
  \hskip -1.0cm
  \includegraphics[height=\columnwidth, angle=-90]{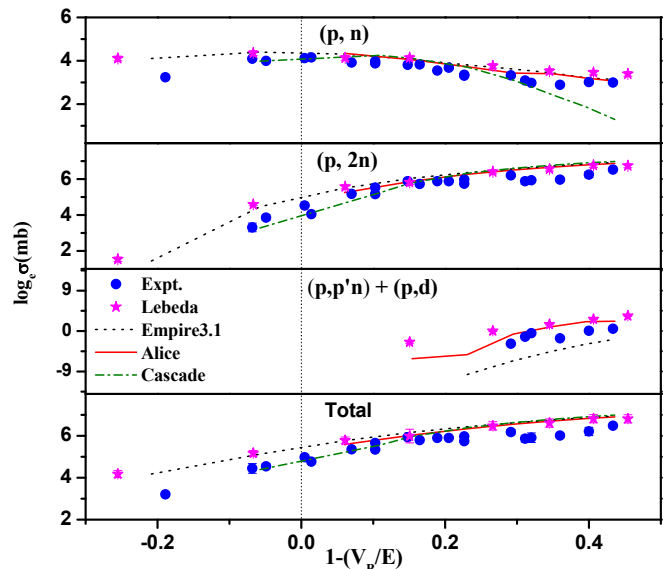}
  %\vskip 1.5cm
  \caption{(Colour online)Individual CS for different evaporation residues and total CS are plotted in logarithmic scale. Legends used are same as in FIG.~\ref{fig5-cs}.}
  \label{fig6-Lncs}
  \end{center}
  \end{figure}

The $\gamma$ rays, used in the CS calculation, were chosen in such a way that they have significant abundance and appear prominently in the spectrum. A representative $\gamma$ spectrum obtained from the irradiated target  along with its catcher foil is shown in FIG.~\ref{fig2-totalspec}.
Most of the $\gamma$ rays were identified to be originated from the decay of $^{150}$Pm, $^{149}$Pm and $^{149}$Nd nuclei. The $^{149}$Pm and $^{149}$Nd nuclei were found to be produced only above E$_p$=7 MeV and E$_p$=11.5 MeV, respectively. The $\gamma$ lines coming from the target impurities have been observed to be very negligible. Some $\gamma$ transitions were found to be present due to the activated products of Al-foils used in the experiment. The 1165.75, 285.95 and 114.31 keV transitions were used for calculating the activity of $^{150}$Pm, $^{149}$Pm and $^{149}$Nd nuclei, respectively. The activities of $^{150}$Pm and $^{149}$Nd were also cross checked by using the 1324.52 keV and 211.31 keV transitions, respectively. Similarly, the Cu-monitor along with its catcher was counted to obtain the yield of the 1115.55 keV transition. This originates from the decay of $^{65}$Zn and was used for calculating the beam flux utilizing the reaction $^{nat}$Cu(p,xn)$^{65}$Zn as mentioned above. The relevant details of the above four $\gamma$ lines are furnished in table~\ref{gamma}. The abundance values of these $\gamma$ lines were taken from the ENSDF data base~\cite{sm150,sm-nd-149} and corrected for their respective conversion coefficients while using for the activity calculation. The said conversion coefficients have been calculated by using the BrIccv2.3S code avaiable in~\cite{bricc}, using the multipolarities taken from the ENSDF databases~\cite{sm150,sm-nd-149}.
\begin{table*}
\caption{The details of the measured reactions and the characteristic $\gamma$ rays used for the measurements of CS.}
\begin{tabular}{|c|c|c|c|c|c|c|}
\hline
\hline
Reaction&Q value$^{\footnotemark}$&$\gamma$ Energy&Multipolarity&Abundances($\gamma$+CE)&Conv. Coeff&Abundance($\gamma$)\\
&(MeV)&(keV)&&$\%$&&$\%$\\
\hline
&&&&&&\\
$^{150}$Nd(p,n)$^{150}$Pm&-0.865&1165.75&E1&15.8&0.000792&14.64\\
&&&&&&\\
&&1324.51&-&17.5&-&17.5\\
&&&&&&\\
$^{150}$Nd(p,2n)$^{149}$Pm&-6.473&285.95&M1(+E2)&3.4&0.079&3.15\\
&&&&&&\\
$^{150}$Nd(p,p$^{\prime}$n)$^{149}$Nd&-7.38&&&&&\\
&&&&&&\\
&&114.31&M1+E2&40.0&1.24&17.85\\
&&&&&&\\
&&211.31&M1+E2&30.8&0.179&26.12\\
&&&&&&\\
$^{150}$Nd(p,d)$^{149}$Nd&-5.156&&&&&\\
&&&&&&\\
$^{65}$Cu(p,n)$^{65}$Zn&-2.134&1115.5&M1+E2&50.04&0.000194&50.03\\
&&&&&&\\
\hline
\end{tabular}
\label{gamma}
\footnotetext{Q values have been calculated by using the mass defects($\Delta$)taken from Ref.~\cite{wallet}.}
\end{table*}
 The $\log _e area$ for each of these photopeaks was plotted as a function of time elapsed from the EOB and fitted with a linear function. The said plots obtained for a particular proton energy value is represented in FIG.~\ref{fig3-decay}.
 The fitted plot, when extrapolated to t=0 (EOB), yields N$_0$ for a particular evaporation residue at a particular proton energy. The error in N$_0$ was considered to be originated from statistical uncertainty($\le$2$\%$).
The absolute efficiency of the detector was calculated from 53 keV to 1408 keV from the area obtained under the known photopeaks and the known values of the absolute intensities of the $\gamma$ lines of $^{152}$Eu and $^{133}$Ba decay. The obtained efficiency values were plotted as a function of  $\gamma$ energy in a Log-Log scale as shown in FIG.~\ref{fig4-eff}. The error in the efficiency was found to be maximum of 5$\%$ considering the error in the area of weaker photopeaks as well as activity (dps) of the standard sources.
  The efficiencies were fitted with a fourth order polynomial function as given below.
\begin{equation}\label{}
 %\nonumber
   \log_e\epsilon = a_0 + \sum_{n=1-4}a_n(\log_eE_{\gamma})^n
   \nonumber
 \end{equation}
 The efficiency values for the characteristic decay $\gamma$ rays were calculated from the interpolation of this fitted curve, while considering E$_{\gamma}$ in keV..
The CS values for the individual evaporation residues in the present range of proton energy were calculated using the equation~\ref{eqn-cs}. The errors associated with N$_0$($\le$2$\%$), $\epsilon$($\le$5$\%$), N($\le$1.5$\%$) and $\phi$($\le$5$\%$) contributed towards the error in the CS value.
The excitation function for the different evaporation residues are plotted in FIG.~\ref{fig5-cs} along with the total fusion CS calculated by adding the above individual evaporation channels.
In order to exhibit the pattern of the excitation function below  and above the Coulomb barrier (V$_B$), the $\log _e \sigma$ values were plotted against (1-$\frac{V_B}{E_p}$) in FIG.~\ref{fig6-Lncs} considering V$_B$=8.46 MeV.
The experimental CS values for $^{150}$Nd(p,n)$^{150}$Pm reaction, obtained in the present work, exhibit a maximum value of 63.7(4.9)mb at around coulomb barrier (E$_p$ $\sim$ 8.6 MeV). The CS value for (p,n) channel drops down to $\sim$ 25mb at the proton energy below the Coulomb barrier by 1.5 MeV. For the $^{150}$Nd(p,2n)$^{149}$Pm reaction, CS values showed a gradual increase in the present energy range of the proton beam. A similar increase was also observed in case of $^{150}$Nd(p,p$^{\prime}$n)$^{149}$Nd reaction with a threshold energy of 12 MeV. However, the individual values of CS for (p,p$^{\prime}$n) channel are much less than that for (p,xn) channels.

The reaction CS values obtained from the work of Lebeda {\it et al.,}~\cite{lebeda-1,lebeda-2} have been scaled up by taking into account the isotopic enrichment of $^{150}$Nd~\cite{wallet}. These values have been shown in FIG.~\ref{fig5-cs} and FIG.~\ref{fig6-Lncs} in order to compare with the present measurement. These calculated values are higher compared to the data obtained from the present measurement, specially in the higher energy range for the (p, 2n) and (p, p$^{\prime}$n) channels and at lower energy for the (p, n) channel. The CS values obtained in the present work have been tabulated in table~\ref{tab-cs} and have been compared with the data obtained by O. Lebeda {\it et al.}~\cite{lebeda-1,lebeda-2}.
\begin{table*}
\caption{The Reaction CS values obtained in the present work for the reaction $^{150}$Nd(p,xnyp).}
\begin{tabular}{|c|cccccccc|}
\hline
\hline
Proton Energy&\multicolumn{8}{c|}{Reaction CS ($\sigma$)$^{\footnotemark}$}\\
(MeV)&\multicolumn{8}{c|}{mb}\\
%&\multicolumn{8}{c|}{}\\
&\multicolumn{2}{c}{$^{150}$Pm}&\multicolumn{2}{c}{$^{149}$Pm}&\multicolumn{2}{c}{$^{149}$Nd}&\multicolumn{2}{c|}{Total}\\
&\multicolumn{2}{c}{(p,n)}&\multicolumn{2}{c}{(p,2n)}&\multicolumn{2}{c}{(p,p$^{\prime}$n)}&\multicolumn{2}{c|}{(p,xnyp)}\\
&\multicolumn{2}{c}{}&\multicolumn{2}{c}{}&\multicolumn{2}{c}{+}&\multicolumn{2}{c|}{}\\
&\multicolumn{2}{c}{}&\multicolumn{2}{c}{}&\multicolumn{2}{c}{(p,d)}&\multicolumn{2}{c|}{}\\
%&\multicolumn{2}{c}{}&\multicolumn{2}{c}{}&\multicolumn{2}{c}{}&\multicolumn{2}{c|}{}\\
&Present& Lit.&Present&Lit.&Present&Lit.&Present&Lit.\\
&work&~\cite{lebeda-1,lebeda-2}&work&~\cite{lebeda-1,lebeda-2}&work&~\cite{lebeda-1,lebeda-2}&work&~\cite{lebeda-1,lebeda-2}\\
\hline
&&&&&&&&\\
6.74&-&60.66(6.74)&-&4.63(0.59)&-&-&-&65.29(10.99)\\
&&&&&&&&\\
7.12&25.55(1.88)&-&-&-&-&-&25.55(1.88)&64.18(10.8)\\
&&&&&&&&\\
7.92&59.95(4.31)&-&27.36(6.66)&-&-&-&87.31(22.17)&-\\
&&&&&&&&\\
7.93&-&77.86(8.69)&-&98.26(6.39)&-&-&-&176.13(22.75)\\
&&&&&&&&\\
8.06&54.22(4.04)&-&47.23(6.58)&-&-&-&101.45(16.03)&-\\
&&&&&&&&\\
8.49&61.81(4.56)&-&92.57(8.99)&-&-&-&154.38(18.82)&-\\
&&&&&&&&\\
8.58&63.68(4.94)&-&57.12(4.79)&-&-&-&120.80(13.8)&-\\
&&&&&&&&\\
9.01&-&62.43(6.92)&-&262.50(30.15)&-&-&-&324.94(51.86)\\
&&&&&&&&\\
9.10&50.15(3.63)&-&176.16(13.61)&-&-&-&226.31(23.97)&-\\
&&&&&&&&\\
9.42&53.49(3.97)&-&251.74(19.47)&-&-&-&305.22(32.70)&-\\
&&&&&&&&\\
9.43&48.43(3.75)&-&174.56(13.65)&-&-&-&222.98(24.54)&-\\
&&&&&&&&\\
9.93&45.36(3.36)&-&358.17(41.79)&-&-&-&392.85(54.14)&-\\
&&&&&&&&\\
9.96&-&62.79(6.92)&-&336.99(37.25)&-&0.087(0.025)&-&399.87(130.18)\\
&&&&&&&&\\
10.13&46.13(3.56)&-&306.98(22.29)&-&-&-&353.11(37.43)&-\\
&&&&&&&&\\
10.43&34.80(2.68)&-&358.05(40.86)&-&-&-&392.85(54.14)&-\\
&&&&&&&&\\
10.64&39.58(3.10)&-&358.47(25.57)&-&-&-&398.05(42.19)&-\\
&&&&&&&&\\
10.93&27.34(2.09)&-&358.47(25.57)&-&-&-&398.05(42.19)&-\\
&&&&&&&&\\
10.94&28.65(2.07)&-&393.89(32.42)&-&-&-&422.54(46.23)&-\\
&&&&&&&&\\
11.53&-&43.10(4.61)&-&604.82(65.63)&-&1.04(0.13)&-&648.96(130.05)\\
&&&&&&&&\\
11.94&27.76(2.19)&-&495.02(35.47)&-&0.06(0.01)&-&522.85(91.88)&-\\
&&&&&&&&\\
12.28&21.98(1.67)&-&357.57(61.25)&-&0.27(0.02)&-&379.83(76.45)&-\\
&&&&&&&&\\
12.44&19.62(1.50)&-&377.11(27.77)&-&0.62(0.15)&-&397.35(102.97)&-\\
&&&&&&&&\\
12.92&-&34.05(3.72)&-&704.15(78.04)&-&4.43(0.53)&-&742.64(145.99)\\
&&&&&&&&\\
13.21&17.95(0.49)&-&389.38(38.42)&-&0.20(0.003)&-&407.54(42.47)&-\\
&&&&&&&&\\
14.10&20.46(1.60)&-&517.91(86.87)&-&1.05(0.15)&-&539.42(125.25)&-\\
&&&&&&&&\\
14.25&-&31.75(3.37)&-&865.56(95.78)&-&14.08(1.67)&-&911.39(176.55)\\
&&&&&&&&\\
14.94&20.01(1.58)&-&685.26(48.72)&-&1.67(0.12)&-&706.95(91.32)&-\\
&&&&&&&&\\
15.50&-&29.79(3.19)&-&847.82(94.00)&-&29.44(3.55)&-&907.06(177.49)\\
&&&&&&&&\\
\hline
\end{tabular}
\footnotetext{The CS values given as Lit. value has been calculated by multiplying the CS obtained from the work of O. Lebeda {\it et al.,} with the isotopic enrichment of $^{150}$Nd~\cite{wallet}.}
\label{tab-cs}
\end{table*}

\section{Nuclear Model Calculation}
\label{theory-cs}
The experimental excitation functions were corroborated using the statistical model codes CASCADE~\cite{cascade}, ALICE~\cite{alice} and EMPIRE3.1~\cite{empire}.
All the calculations could provide an accurate overall description for all the excitation functions, however, varied in the prediction of the exact position of the maximum for the (p, n) channel as well as the ascending or descending slopes. A variation between the experimental data and the theoretical prediction was observed for the (p, 2n) channel after E$_p$ = 12 MeV. However, this deviation may be accounted considering the multitude of uncertainties and the limitations embedded in the theoretical codes. In the following subsections, the relevant details on each of the model calculations have been furnished.

\subsection{CASCADE}

The calculation with CASCADE code assumes the formation of a compound nucleus in statistical equilibrium and the intensities of different evaporation residues are calculated applying Hauser-Feshbach formula in combination with the statistical nuclear model. In the present work, the fusion CS was estimated considering the diffuseness parameter $\delta$l = 0.5. The optical model potential parameters of Willmore and Hodgson were used for neutron transmission coefficients while the parameters of Perey were used for proton transmission coefficients~\cite{perey}. In order to check the dependence on the transmission coefficients, the calculations were also performed considering the optical model potential parameters of Becchetti and Greenless~\cite{becchetti} for neutron and proton decay. The results were very similar as obtained using the previous potentials.  The $\gamma$ decay was also considered in the calculation and was taken as 0.3 of the Weiskoff unit for E1 decay. The level density prescription of Reisdorf~\cite{reisdorf} was adopted for the calculations. The (p,n) and (p,2n) CS values obtained using CASCADE are shown in FIG.~\ref{fig5-cs} and~\ref{fig6-Lncs} with dot-dashed line. The (p,n) and (p,2n) CS are quite well reproduced except at higher energies where the calculation underestimates the (p,n) CS whereas overestimates the (p,2n) CS. No CS was obtained for (p,p$^{\prime}$n) channel for all the incident proton energy. This could be due to the inherent problem of the CASCADE code in predicting low CS, as similar discrepancy is observed for (p,n) channel at higher proton energy in comparison to other models.

\subsection{ALICE}

The Alice/91~\cite{alice} is a precompound and evaporation model code system for calculating the excitation functions and angular distribution of emitted particles in nuclear reactions. The model performs several types of calculations and combinations including a standard Weisskopf-Ewing evaporation~\cite{weisskopf} with multiple particle emission, s-wave approximation to give an upper limit to the enhancement of $\gamma$ ray deexcitation due to angular momentum effects and an evaporation calculation that can include fission competition via the Bohr-Wheeler approach. ALICE91 calculates precompound decay via Hybrid and geometry dependant hybrid model (GDH)~\cite{gdh} with multiple precompound decay algorithms, single and double differential spectra, and reaction product CS. In this code, the beginning of any particle induced nuclear reaction is characterized by the configuration of the initially excited number of particles and holes, called as excitons. The excitons are described with respect to the ground state configuration of the compound system. The intermediate state of the system is defined by the excitation energy and the numbers of excitons.

The binding energies and Q-values were calculated by using the database of the experimental masses by Wapstra and Audi~\cite{wapstra} wherever available and calculated from Myers and Swiatecki mass formula~\cite{myers}, otherwise. The inverse CS were calculated using the Optical Model subroutine with the optical model parameters of Becchetti and Greenlees~\cite{becchetti}. The level densities were calculated using the Ignatyuk's formula~\cite{ignatyuk} with the level density parameter a = A/9 MeV$^{-1}$, as the default option of the code. The present calculation reproduced the experimental data points quite well. However, the overprediction of the higher energy data points could be explained by the uncertainties in some of the parameters used in the code.

\subsection{EMPIRE}

The nuclear reaction code EMPIRE  (version 3.1 Rivoli), developed by Herman {\it et al.}~\cite{empire}, is a modular system of nuclear reaction codes. It comprises of different nuclear models and is designed to perform nuclear reaction calculations over a wide range of incident energies and projectiles. The code makes use of an improved version of the Hauser-Feshbach theory for the statistical part and the exciton model for the preequilibrium part of a nuclear reaction. Different input parameters like nuclear masses, ground state deformations, discrete levels, $\gamma$ ray strength functions, etc. were retrieved from the standard library RIPL-3~\cite{ripl} included in the code. The particle transmission coefficients were calculated using the optical model routine ECIS06~\cite{ecis}. The preequilibrium reactions with angular momentum conservation were considered using the code PCROSS , which can calculate nucleons, clusters and $\gamma$ emission spectra in terms of the exciton model, based on the Iwamoto-Harada model~\cite{iwamoto}. The width fluctuation correction based on the Hofmann, Richert, Tepel and Weidenmuller (HRTW) model~\cite{hrtw} was found to have almost no effect in the present calculations. The microscopic Hartree-Fock-Bogoliubov (HFB) level densities were taken from an internal file included in RIPL-3. In the present energy range, the calculation slightly overpredicts the experimental results in case of neutron evaporation reactions. However, the overall trend follows the experimental data points.

\section{Discussion}
\label{dis}

The excitation functions for the p-induced reaction on the 97.65$\%$ enriched $^{150}$Nd target were measured by $\gamma$-decay spectroscopy following the stacked-foil activation technique.
In the present work, the absolute CS for the reaction $^{150}$Nd(p, xnyp) has been measured for the first time, following the use of the enriched $^{150}$Nd target. The present measurement also yields the first set of data points for these reaction CS at several proton energy values covering an energy range of 7 - 15 MeV.
In this work all sorts of precautions were taken to avoid the over or under estimation of reaction CS. Apart from the use of an enriched target, the target enrichment as well as the Nd:O ratio in the prepared target was considered for the calculation of the number of target Nd atoms. The present work also involved the use of separate Al-catcher foils of appropriate thickness corresponding to each target and monitor in order to ensure the complete collection of the recoiling evaporation residues. Most importantly, the dead time of the detector was kept lower ($\le$10$\%$) in the present work than that ($\le$40$\%$) allowed in the work of Lebeda {\it et al.}~\cite{lebeda-1,lebeda-2}. The abundances used for the $\gamma$ transitions were also corrected for the appropriate conversion as per their multipolarity. Wherever possible, the obtained CS values were cross checked by using the activities from two different $\gamma$ transitions.
The CS values at several proton energies were calculated from two different measurements by making appropriate target stacks as explained in section~\ref{expt} in order to ensure the correctness of our experimental data in the present work.

The excitation functions for the neutron evaporation residues, obtained from the present work, were found to follow the expected trend in the present range of proton energy. The $^{149}$Nd was populated through both (p, p$^{\prime}$n) and (p, d) channels whose individual contribution could not be deciphered in the present experiment. Similarly, the production of $^{149}$Pm has a contribution from the decay of $^{149}$Nd. However, this has been observed to be very negligible compared to the production of $^{149}$Pm from the (p, 2n) reaction channel. The above observation is further supported by the low CS values obtained for $^{149}$Nd in the present work. The theoretical calculations for the excitation functions have been performed with different statistical model codes, {\it viz.,} CASCADE, ALICE and EMPIRE3.1. Both the experimental data and theoretical calculations exhibited reasonably similar trend in the entire proton energy range.

The CS values, obtained from the work of Lebeda {\it et al.,}~\cite{lebeda-1,lebeda-2}, were scaled up to the absolute values by considering the isotopic enrichment ratio of Nd. These reconstructed values are higher compared to the absolute CS measured in the present work. It is obvious that both the absolute CS values directly obtained from the present experiment and the CS values reconstructed from the work by Lebeda {\it et al.,} show a reasonable similarity to that predicted by the theoretical models. However, the extent of closeness between these two sets of experimental CS values with the theoretical predictions is different. This might be due to the combination of several unknown experimental conditions involved in a multi-step measurement procedure. Again, the theoretical calculations, involving hosts of uncertainties in the large number of input parameters, are necessarily capable of reproducing the overall trend of the excitation function.

\section{Acknowledgement}
\label{ack}

The effort of the staffs and members of the K=130 cyclotron operation group at VECC, Kolkata, is gratefully acknowledged for providing high quality stable proton beam. The authors would like to acknowledge the effort of Dr. P. Acharya, Radio Chemistry Division, BARC for carrying out the neutron activation analysis of the targets used in the experiment. A. Saha is grateful for his UGC Fellowship No:  . The efforts of Mr. R. K. Chatterjee, RCD, VECC is acknowledged for preparation of the targets.

\end{document}